\documentclass[%
 aip,
 jcp,%
 amsmath,amssymb,
 reprint,%
]{revtex4-1}

\usepackage{graphicx}
\usepackage{dcolumn}
\usepackage{bm}
\usepackage{color} 
\usepackage{amsmath,amssymb}
\usepackage{epsfig}
\usepackage{bbold}
\usepackage{url}
\usepackage{lineno}
\usepackage{amsmath,amssymb}
\usepackage[colorlinks=true,citecolor={blue},urlcolor={red},linkcolor={blue}]{hyperref}
\usepackage{ulem}
\begin{document}
\title{Time-dependent optimized coupled-cluster
method for multielectron dynamics III: A second-order many-body perturbation approximation}
\author{Himadri Pathak}%
\email{pathak@atto.t.u-tokyo.ac.jp}
\affiliation{Department of Nuclear Engineering and Management, School of Engineering, The University of Tokyo, 7-3-1 Hongo, Bunkyo-ku, Tokyo 113-8656, Japan}%
\author{Takeshi Sato}
\email{sato@atto.t.u-tokyo.ac.jp}
\affiliation{Department of Nuclear Engineering and Management, School of Engineering, The University of Tokyo, 7-3-1 Hongo, Bunkyo-ku, Tokyo 113-8656, Japan}
\affiliation{Photon Science Center, School of Engineering, The University of Tokyo, 7-3-1 Hongo, Bunkyo-ku, Tokyo 113-8656, Japan}
\affiliation{Research Institute for Photon Science and Laser Technology, The University of Tokyo, 7-3-1 Hongo, Bunkyo-ku, Tokyo 113-0033, Japan}
\author{Kenichi L. Ishikawa}
\email{ishiken@n.t.u-tokyo.ac.jp}
\affiliation{Department of Nuclear Engineering and Management, School of Engineering, The University of Tokyo, 7-3-1 Hongo, Bunkyo-ku, Tokyo 113-8656, Japan}
\affiliation{Photon Science Center, School of Engineering, The University of Tokyo, 7-3-1 Hongo, Bunkyo-ku, Tokyo 113-8656, Japan}
\affiliation{Research Institute for Photon Science and Laser Technology, The University of Tokyo, 7-3-1 Hongo, Bunkyo-ku, Tokyo 113-0033, Japan}
\begin{abstract}
We report successful implementation of the time-dependent second-order many-body perturbation theory
using optimized orthonormal orbital functions called
time-dependent optimized second-order many-body perturbation theory [TD-OMP2] to reach out 
to relatively larger chemical systems for the study of intense-laser-driven multielectron dynamics.
We apply this method to strong-field ionization and high-order harmonic generation (HHG) 
of Ar.
The calculation results are benchmarked against {\color{black}\it ab initio} time-dependent
complete-active-space self-consistent field (TD-CASSCF), time-dependent optimized coupled-cluster double (TD-OCCD), and
time-dependent Hartree-Fock (TDHF) methods, {\color{black}as well as a single active electron (SAE) model}
to explore the role of electron correlation.
\end{abstract}
\date{\today}
\maketitle
\section{Introduction}
Laser-driven multielectron dynamics has become an active area of research, thanks to the remarkable advance
in laser technologies, which has made it possible to measure and control the electronic motion
\cite{corkum2007attosecond, krausz2009attosecond, itatani2004, goulielmakis2010real, sansone2010electron,
schultze2010, klunder2011probing, belshaw2012observation, calegari2014, smirnova2009high}.
Atoms and molecules interacting with laser pulses of intensity $\gtrsim 10^{14}\,{\rm W/cm}^2$ in the visible to mid-infrared spectral range, exhibit highly, even nonperturbatively nonlinear response such as above-threshold ionization (ATI), tunneling ionization, nonsequential double ionization (NSDI), and high-order harmonic generation (HHG).
The HHG process is one of the key elements in the study of light-matter interaction in the attosecond time-scale, delivering
ultrashort coherent light pulses in the extreme-ultraviolet (XUV) to the soft x-ray regions, which carry the information on the electronic structure and dynamics of the generating medium. \cite{antoine1996attosecond}
The HHG spectra are characterized by a plateau where the intensity of the emitted radiation remains nearly constant up to many orders,
followed by an abrupt cutoff.\par
%
In principle, the multielectron dynamics and electron correlation \cite{ishikawa2015review, tikhomirov2017high, likumar, PhysRevLett.santra} are exactly described by the time-dependent Schr{\"o}dinger equation (TDSE). 
However, direct numerical integration of TDSE is not feasible for systems with more than two electrons \cite{parker1998intense, parker2000time, pindzola1998time, laulan2003correlation, ishikawa2005above,
feist2009probing, ishikawa2012competition, sukiasyan2012attosecond, vanroose2006double, horner2008classical}.
As a result, single-active-electron (SAE) approximations has been widely
used \cite{krause1992jl, kulander1987time}, in which only the outermost electron is explicitly treated under the effect of the other electrons modeled by an effective potential.
Whereas SAE has been useful in numerically exploring different strong-field phenomena, the electron correlation is missing in this approximation. \par
Therefore, various tractable {\it ab initio} methods have actively been developed for theoretical description of correlated multielectron dynamics in intense laser fields.
{\color{black}Among the most reliable approaches to serve the purpose are the multiconfiguration time-dependent Hartree-Fock (MCTDHF) method
\cite{caillat2005correlated, kato2004time, nest2005multiconfiguration, haxton2011multiconfiguration, hochstuhl2011two} and time-dependent complete-active-space self-consistent-field (TD-CASSCF)
method \cite{sato2013time, sato2016time}. 
In MCTDHF, the electronic wavefunction is expressed in terms of full configuration interaction (FCI) expansion,
$\Psi(t)=\sum_{\bm I}C_I(t)\Phi_{\bm I}(t)$, where both CI coefficients $\{C_{\bm I}(t)\}$ and {\it occupied} spin-orbitals $\{\psi_p(t)\}$ constituting Slater determinants $\{\Phi_{\bm I}(t)\}$ are propagated in time. The TD-CASSCF method
flexibly classifies occupied orbital space into {\it frozen-core} (doubly occupied and fixed in time),
{\it dynamical-core} (doubly occupied but propagated in time) and {\it active} (fully correlated and propagated in time) subspaces. }
Though accurate and powerful, the computational cost of the MCTDHF and TD-CASSCF methods scales factorially with the number of correlated electrons.

More approximate and thus less demanding methods such as the time-dependent restricted-active-space self-consistent field (TD-RASSCF),\cite{miyagi2013time, miyagi2014time, haxton2015two}
and time-dependent occupation-restricted multiple-active-space (TD-ORMAS) \cite{sato2015time} have been developed by further flexibly classifying {\it active} orbital sub-space
to target larger chemical systems by limiting CI expansion of the
wavefunction up to a manageable level.
The TD-RASSCF and TD-ORMAS methods achieve a polynomial, instead of
factorial, cost scaling, and state-of-the-art real-space implementations
have turned out to be of great utility.
\cite{haxton2015two,sato2016time,Omiste:2017,orimo2018implementation} 
{\color{black}
(See Ref.~\citenum{ishikawa2015review} for a review of {\it ab initio} wavefunction-based
methods for multielectron dynamics, Ref.~\citenum{anzaki:2017} and
references therein for extension to correlated electron-nuclear
dynamics, and Ref.~\citenum{Lode:2020} for a perspective on multiconfiguration approaches
for indistinguishable particles.)}
Nevertheless, truncated-CI-based methods, even with time-dependent orbitals, have a general drawback of not being size extensive.

To regain the size-extensivity at the same level of truncation, we have recently derived and numerically implemented
a time-dependent optimized coupled-cluster (TD-OCC) method \cite{sato2018communication}
using optimized orthonormal orbitals 
where both orbitals
and amplitudes
are time-dependent.
Our TD-OCC method is the time-dependent formulation of the stationary orbital optimized coupled-cluster method. \cite{sherrill1998energies, krylov1998size} 
{\color{black}
We have implemented the TD-OCC method with up to triple excitation
amplitudes (TD-OCCD and TD-OCCDT), and applied it to multielectron dynamics in Ar induced by a strong laser pulse to obtain a good agreement
with the fully-correlated TD-CASSCF method within the same active
orbital space.}
In earlier work, Kvaal reported \cite{kvaal2012ab} an orbital adaptive
coupled-cluster (OATDCC) method built upon the work of Arponen using biorthogonal orbitals; \cite{arponen1983variational}
though, applications to laser-driven dynamics have not been addressed.
The polynomial scaling TD-OCC \cite{sato2018communication} or the OATDCC \cite{kvaal2012ab} method can reduce the computational
cost to a large extent in comparison to the general factorially scaling MCTDHF methods.

{\color{black}As a cost-effective approximation of the TD-OCCD method,
we have recently developed a method called
TD-OCEPA0\cite{pathak2020timedependent} based on the simplest version of the coupled-electron pair
approximation,\cite{meyer1971ionization, ahlrichs1985coupled,wennmohs2008comparative,
neese2009efficient,kollmar2010coupled,malrieu2010ability} or
equivalently the linearized CCD
approximation,\cite{vcivzek1966correlation} popular in quantum
chemistry. 
The computational cost of the TD-OCEPA0 method scales as 
$N^6$, with $N$ being the number of active orbitals.
Although this is the same scaling as that of the parent TD-OCCD method, TD-OCEPA0 is much
more efficient than TD-OCCD due to the linearity of amplitude equations and
avoidance of a separate solution for the de-excitation amplitudes,
resulting from the Hermitian structure of the underlying Lagrangian.\cite{pathak2020timedependent}}

To enhance the applicability to even larger chemical systems, we are
looking for further approximate versions {\color{black}with a lower
computational scaling} in the TD-OCC framework.
The coupled-cluster method is intricately connected with the many-body perturbation theory.
It allows one to obtain finite-order perturbation theory energies and wavefunction from the coupled-cluster equations. \cite{bartlettreviewmbptcc}
The computation of the second-order energy requires the first-order wavefunction, and only doubly excited determinants
have contributions in the first-order correction to the reference wavefunction.
Thus, second-order many-body perturbation theory {\color{black}(MP2)} can
be seen as an approximation to the coupled-cluster double (CCD) method,
{\color{black} having a lower $N^5$ scaling.}

The MP2 method with optimized orbitals has been developed by Bozkaya {\it et al.} \cite{bozkaya2011quadratically}
for the stationary electronic structure calculations by 
minimization of the {\color{black}so-called} MP2-$\Lambda$ 
functional.
In earlier work, the optimization of the MP2 energy was based on the minimization of the Hylleraas
functional with respect to the orbital rotation. \cite{adamowicz1987optimized, neese2009assessment} While both of these techniques provide identical
energy at the stationary point,
{\color{black}the $\Lambda$ functional-based derivation has an advantage that it can
be easily extended 
to higher-order many-body perturbation theory.\cite{bozkaya2011orbital}}

In this {\color{black}article}, we {\color{black}propose} time-dependent orbital optimized second-order many-body perturbation
theory (TD-OMP2) {\color{black}as an approximation to the TD-OCCD method, based on the
time-dependent (quasi) variational principle.
The TD-OMP2 method inherits important features of size
extensivity and gauge invariance from TD-OCC, with a reduced computational
cost of $O(N^5)$.}
As a first test case, we {\color{black}apply the TD-OMP2} method to
{\color{black}electron dynamics} in Ar atom {irradiated by a strong laser field}, and
{\color{black} compare the results }with those computed at {\color{black}SAE,} TDHF, TD-OCCD, and TD-CASSCF levels to understand where really TD-OMP2 stands in describing the effects of correlation in the laser-driven multielectron dynamics.
{\color{black}
It should be emphasized that the present TD-OMP2 method, albeit named after a perturbation theory, can be applied to laser-induced, nonperturbative
electron dynamics, since the laser-electron interaction is
fully (nonperturbatively) included in the zeroth-order description with
time-dependent orbitals.}

This paper is organized as follows. Our formulation of the TD-OMP2 method is presented in Sec. \ref{sec2}.
Numerical applications are described and discussed in Sec. \ref{sec3}. The concluding remarks are given in Sec. \ref{sec4}.
Hartree atomic units are used unless otherwise stated{\color{black} , and Einstein convention is implied throughout for summation over orbital indices.}

\begin{table*}[ht!]
\caption{\label{tab:comparison} Comparison of ground state energy (in Hartree) of BH molecule in DZP basis with PSI4\cite{psi4} program package.}
\begin{ruledtabular}
\begin{center}
\begin{tabular}{lrrrrrrr}
\multicolumn{1}{c}{Bond Length} &
\multicolumn{2}{c}{MP2} & &
\multicolumn{2}{c}{OMP2}  \\
\cline{2-3}
\cline{5-6}
\\
(bohr)& this work & PSI4\cite{psi4} & &this work & PSI4\cite{psi4} \\
\hline\\
1.8&$-$25.149288192&$-$25.149288192&&$-$25.149565428&$-$25.149565428\\
2.0&$-$25.183068430&$-$25.183068429&&$-$25.183367319&$-$25.183367319\\
2.2&$-$25.196855310&$-$25.196855310&&$-$25.197186611&$-$25.197186611\\
2.4(r$_e$)&$-$25.198570797&$-$25.198570797&&$-$25.198947432&$-$25.198947432\\
2.8&$-$25.183573786&$-$25.183573786&&$-$25.184093435&$-$25.184093435\\
3.2&$-$25.159043339&$-$25.159043339&&$-$25.159809546&$-$25.159809546\\
3.6&$-$25.133018128&$-$25.133018128&&$-$25.134185728&$-$25.134185728\\
4.0&$-$25.108605403&$-$25.108605402&&$-$25.110381925&$-$25.110381924\\
5.0&$-$25.059981388&$-$25.059981388&&$-$25.064548176&$-$25.064548176\\
6.0&$-$25.029750598&$-$25.029750598&&$-$25.039814328 &$-$25.039814327\\
7.0&$-$25.016553779&$-$25.016553780&&$-$25.037103689&$-$25.037103689\\
\end{tabular}
\end{center}
\end{ruledtabular}
{Gaussian09 program\cite{gaussian09} is used to generate the required one-electron, two-electron and the overlap integrals,
required for the imaginary time propagation of EOMs in the orthonormalized gaussian basis. 
A convergence cut off of 10$^{-15}$ Hartree of energy difference is chosen in subsequent time steps.}
\end{table*}
\section{TD-OMP2 method}\label{sec2}
{\color{black}
Let us consider the electronic Hamiltonian $H$ of the following form, 
\begin{eqnarray}
H &=& \sum_i^{\color{black}N_e} h({\bm r}_i,{\bm p}_i)+\sum_{i<j}^{\color{black}N_e}\frac{1}{\lvert{\bm{r}_i-\bm{r}_j}\rvert},
\end{eqnarray}
\begin{eqnarray}
h({\bm r},{\bm p}) &=& h_0({\bm r},{\bm p}) + V_{\text{ext}}({\bm r},{\bm p},t),
\end{eqnarray}
where {\color{black}$N_e$ is the number of electrons,} $\bm r_i$ and $\bm p_i$ are the position and canonical momentum, respectively, of
electron $i$, $h_0$ is the field-free one-electron Hamiltonian, 
and $V_{\text{ext}}$ is the laser-electron interaction.}
The Hamiltonian in the second-quantization notation can be written as,
\begin{eqnarray}
\hat{H}
&=& \hat{h} + \hat{v}, 
\end{eqnarray}
\begin{eqnarray}
\hat{h} = h^\mu_\nu \hat{E}^\mu_\nu, \hspace{1em}
\hat{v} = \frac{1}{2}u^{\mu\gamma}_{\nu\lambda} \hat{E}^{\mu\gamma}_{\nu\lambda}
 = \frac{1}{4}v^{\mu\gamma}_{\nu\lambda} \hat{E}^{\mu\gamma}_{\nu\lambda},
\end{eqnarray}
where $\hat{E}^\mu_\nu = \hat{c}^\dagger_\mu\hat{c}_\nu$, 
$\hat{E}^{\mu\gamma}_{\nu\lambda} =
\hat{c}^\dagger_\mu\hat{c}^\dagger_\gamma\hat{c}_\lambda\hat{c}_\nu$, 
{\color{black} and
$\hat c_\mu^\dag (\hat{c}_\mu)$ is the creation (annihilation) operator for
the set of orthonormal $2n_{\rm bas}$ spin-orbitals $\{\psi_\mu\}$,
with $n_{\rm bas}$ being the number of basis functions (or grid points)
to represent the spatial part of $\psi_\mu$}. The operator 
matrix elements $h^\mu_\nu$, $u^{\mu\gamma}_{\nu\lambda}$, and $v^{\mu\gamma}_{\nu\lambda}$ are defined as
\begin{eqnarray}\label{eq:int1}
h^\mu_\nu = \int dx_1 \psi^*_\mu(x_1) h(\bm{r}_1,\bm{p}_1) \psi_\nu(x_1),
\end{eqnarray}
\begin{eqnarray}\label{eq:int2}
 u^{\mu\gamma}_{\nu\lambda} = \int\int dx_1dx_2
 \frac{\psi^*_\mu(x_1)\psi^*_\gamma(x_2)\psi_\nu(x_1)\psi_\lambda(x_2)}{|\bm{r}_1-\bm{r}_2|},
\end{eqnarray}
and $v^{\mu\gamma}_{\nu\lambda} = u^{\mu\gamma}_{\nu\lambda}-u^{\mu\gamma}_{\lambda\nu}$, where $x_i=(\bm{r}_i,\sigma_i)$ is a spatial-spin coordinate.
%
%
{\color{black}
The complete set of $2n_{\rm bas}$ spin-orbitals (labeled with
$\mu,\nu,\gamma,\lambda$) is divided into $n_{\rm occ}$ {\it occupied} ($o,p,q,r,s$) and
$2n_{\rm bas}-n_{\rm occ}$ {\it virtual} spin-orbitals having
nonzero and vanishing occupations, respectively, in the
total wavefunction. The occupied spin-orbitals
are classified into $n_{\rm core}$ {\it core} spin-orbitals 
which are occupied in the reference $\Phi$ and kept uncorrelated, and
$n_{\rm act}=n_{\rm occ}-n_{\rm core}$ {\it active} spin-orbitals ($t,u,v,w$) among which the
$N_{\rm act}=N_e-n_{\rm core}$ active electrons are correlated. The active
spin-orbitals are further split into those in the {\it hole} space
($i,j,k,l$) and the {\it particle} space ($a,b,c,d$), which are defined as those occupied and
unoccupied, respectively, in the reference $\Phi$.
The core spin-orbitals can further be split into {\it frozen-core} space ($i^{\prime\prime},j^{\prime\prime}$)
and the {\it dynamical-core} space ($i^\prime,j^\prime$).
The frozen-core orbitals are fixed in time, whereas dynamical core orbitals are propagated in time. \cite{sato2013time}
(See Fig.~1 of Ref.~\citenum{sato2018communication} for a pictorial illustration of the orbital subspacing.)}

\subsection{Review of TD-OCCD}

The stationary MP2 method can be considered as an
approximation of the CCD method, where the full CCD energy functional, 
\begin{eqnarray}\label{eq:ccdene}
E&=&\langle \Phi|(1+\hat \Lambda_2)e^{-\hat T_2}\hat He^{\hat T_2}|\Phi\rangle
\end{eqnarray}
is approximated by retaining the terms giving the second-order correction to the reference energy $E_0=\langle\Phi|\hat H|\Phi\rangle$.
Here, $\hat{T}_2=\tau^{ab}_{ij}\hat{E}^{ab}_{ij}$ 
and $\hat{\Lambda}_2=\lambda_{ab}^{ij}\hat{E}_{ab}^{ij}$,
with $\tau^{ab}_{ij}$ and $\lambda_{ab}^{ij}$ being the excitation and de-excitation amplitudes, respectively.
Therefore, we start with the time-dependent CCD Lagrangian of the following form,  
\begin{eqnarray}\label{eq:ccdlag}
L(t)&=&\langle \Phi|(1+\hat \Lambda_2)e^{-\hat T_2}(\hat H-i\partial_t)e^{\hat T_2}|\Phi\rangle,
\end{eqnarray}
which is a natural time-dependent extension of the energy functional, Eq.~(\ref{eq:ccdene}). 
Following Ref.~\citenum{sato2018communication}, we consider 
the real-valued action functional,
\begin{eqnarray}
S &=& \label{eq:action}
\Re
\int_{t_0}^{t_1} L(t)dt = \frac{1}{2} \int_{t_0}^{t_1}\left\{L(t) + L^*(t)\right\} dt,
\end{eqnarray}
and make it stationary, $\delta S=0$, with respect to
the variation of amplitudes $\delta\tau^{ab}_{ij}$,
$\delta\lambda_{ab}^{ij}$ and variations of orthonormality-conserving orbitals
$\delta\psi_\nu$.
The equations of motion (EOMs) for the amplitudes are obtained as
\begin{eqnarray}
i\dot{\tau}^{ab}_{ij}&=& \label{eq:td-ccd_t}
\langle\Phi_{ij}^{ab}|e^{-\hat{T}_2}(\hat{H}-i\hat{X})e^{\hat
 T_2}|\Phi\rangle,\\ 
-i\dot{\lambda}^{ij}_{ab}&=& \label{eq:td-occd_l}
\langle\Phi|(1+\hat
 \Lambda_2)e^{-\hat{T}_2}(\hat{H}-i\hat{X})e^{\hat T_2}|\Phi_{ij}^{ab}\rangle, 
%
%
\end{eqnarray}
where 
$\hat{X}=X^\mu_\nu\hat c^\dag_\mu \hat c_\nu$ with
$X^\mu_\nu=\langle\psi_\mu|\dot{\psi}_\nu\rangle$, 
and those for orbitals as,
\begin{eqnarray}
i|\dot{\psi_p}\rangle &=& \label{orbeom1}
(1-\hat{P}) \hat{F}_p|\psi_p\rangle + |\psi_q\rangle X^q_p,
\end{eqnarray}
\begin{eqnarray}\label{orbeom2}
i\left\{(\delta^a_b\rho^j_i-\rho^a_b\delta^j_i)X^b_j\right\} =
 F^a_j\rho^j_i - \rho^a_bF^{i*}_b,
\end{eqnarray}
where $\hat{P}=\sum_p|\psi_p\rangle\langle\psi_p|$, $F_q^p=\langle
\phi_p|\hat{F}_q|\phi_q\rangle$, with
\begin{eqnarray}
\hat{F}_p|\psi_p\rangle &=& 
\hat{h} |\psi_p\rangle +  \hat{W}^r_s|\psi_q\rangle \rho^{qs}_{or}{(\rho^{-1})}_p^o,
\end{eqnarray}
\begin{eqnarray}
W^r_s(\bm{x}_1)=
\int d\bm{x}_2
\frac{\psi^*_r(\bm{x}_2)\psi_s(\bm{x}_2)}{|\bm{r}_1-\bm{r}_2|}, 
\end{eqnarray}
where $\rho_q^p$ and $\rho_{qs}^{pr}$ are the one- and two-body reduced
density matrices, respectively, defined as
\begin{eqnarray}
\rho^q_p&=&\label{eq:1rdm}
\langle\Phi|(1+\hat{\Lambda_2})e^{-\hat{T_2}}\hat{E}^p_qe^{\hat{T_2}}|\Phi\rangle, \\
\rho^{qs}_{pr}&=&\label{eq:2rdm}
\langle\Phi|(1+\hat{\Lambda_2})e^{-\hat{T_2}}\hat{E}^{pr}_{qs}e^{\hat{T_2}}|\Phi\rangle.
\end{eqnarray}

\subsection{Derivation of TD-OMP2 as an approximation to TD-OCCD}
Now we derive the TD-OMP2 method as an approximation to 
the TD-OCCD method, based on the partitioning of the electronic Hamiltonian, 
\begin{subequations} \label{eqs:part}
\begin{eqnarray}
\hat{H}=\hat{H}^{(0)}+\hat{H}^{(1)},
\end{eqnarray}into the zeroth-order part $\hat{H}^{(0)} = \hat{f} = f^p_q \hat{E}^p_q$ and the perturbation $\hat{H}^{(1)}=\hat{H}-\hat{H}^{(0)}$, with
\begin{eqnarray}
f^p_q=h^p_q+v_{q j}^{p j}
=(h_0)^p_q+v_{q j}^{p j} + (V_{\text{ext}})^p_q,
\end{eqnarray}
\end{subequations}
where $(h_0)^\mu_\nu$ and $(V_{\text{exe}})^\mu_\nu$ are the matrix elements of $\hat{h}_0$ and $\hat{V}_{\textrm{ext}}$, respectively.

Based on this partitioning, we apply the 
Baker-Campbell-Hausdorff expansion to
TD-OCCD Lagrangian of Eq.~(\ref{eq:ccdlag}), 
and retain those terms up to quadratic in $v$, $\tau$, and $\lambda$ 
(thus, contributing to first- and second-order corrections to Lagrangian) to obtain
\begin{eqnarray}\label{eq:td-omp2_lag}
L &=&  L_0 -i\lambda^{ij}_{ab}\dot{\tau}^{ab}_{ij} +
\langle\Phi|\hat{\Lambda}_2(\hat{H}-i\hat{X})|\Phi\rangle \\
&+&
\langle\Phi|[\hat{H}-i\hat{X},\hat{T}_2]|\Phi\rangle +
\langle\Phi|\hat{\Lambda}_2[\hat{f}-i\hat{X},\hat{T}_2]|\Phi\rangle, \nonumber
\end{eqnarray}
where $L_0 = \langle\Phi|(\hat{H}-i\hat{X})|\Phi\rangle$ is the
reference contribution.
Inserting this TD-OMP2 Lagrangian into Eq.~(\ref{eq:action}) and 
making it stationary with respect to amplitude variations derives TD-OMP2 amplitude equations, 
\begin{eqnarray}
i\dot{\tau}^{ab}_{ij}&=&\label{t2eqn1}
v_{ij}^{ab}-p(ij) \bar{f}_j^k
 \tau_{ik}^{ab}+p(ab)\bar{f}_c^a \tau_{ij}^{cb}, \\
-i\dot{\lambda}^{ij}_{ab}\label{lambda2eqn}
&=&v_{ab}^{ij}-p(ij) \bar{f}_k^i
\lambda_{ab}^{kj}+p(ab)\bar{f}_a^c\lambda_{cb}^{ij},
\end{eqnarray}
\begin{eqnarray}\label{eq:fockix}
\bar{f}^p_q=f^p_q-iX^p_q,
\end{eqnarray}
where $p(\mu\nu)$ is the cyclic permutation operator.
Importantly, Eqs.~(\ref{t2eqn1}) and (\ref{lambda2eqn})
reveals that the EOM for $\lambda_{ab}^{ij}$ is just complex conjugate
of that for $\tau_{ij}^{ab}$, resulting in $\lambda^{ij}_{ab}=\tau^{ab*}_{ij}$. 
Therefore, we do not need a separate solution for the {\color{black}$\Lambda_2$ amplitudes}.

We also make the action stationary with respect to the orthonormality-conserving 
orbital variation to derive formally the same orbital EOMs as Eqs.~(\ref{orbeom1}) 
and (\ref{orbeom2}), with one-particle reduced density matrices (1RDM) and 
two-particle reduced density matrices (2RDM) given explicitly as
\begin{eqnarray}
\rho^q_p &=& (\rho_0)^q_p + \gamma^q_p, \label{eq:td-omp2_1rdm}\\
\rho^{qs}_{pr}&=& (\rho_0)^{qs}_{pr} + \Gamma^{qs}_{pr} \label{eq:td-omp2_2rdm},
\end{eqnarray}
where
$(\rho_0)^q_p = \delta^q_j\delta^j_p$
and $(\rho_0)^{qs}_{pr} =
\gamma^q_p \delta^s_j \delta^j_r
+\gamma^s_r \delta^q_j\delta^j_p
-\gamma^q_r \delta^s_j\delta^j_p
-\gamma^s_p \delta^q_j\delta^j_r
+\delta^q_j\delta^j_p\delta^s_k\delta^k_r
-\delta^s_j\delta^j_p\delta^q_k\delta^k_r$ are the reference contributions, and
non-zero elements of $\gamma^q_p$ and $\Gamma^{qs}_{pr}$ are given by
\begin{eqnarray}\label{eq:td-omp2_1rdm_corr}
\gamma_j^i=-\frac{1}{2}\lambda_{cb}^{ki}\tau_{kj}^{cb},\hspace{1em}
\gamma_a^b=\frac{1}{2}\lambda_{ca}^{kl}\tau_{kl}^{cb},
\end{eqnarray}
\begin{eqnarray}\label{eq:td-omp2_2rdm_corr}
\Gamma_{ij}^{ab}=\tau_{ij}^{ab}, \hspace{1em}
\Gamma_{ab}^{ij}=\lambda_{ab}^{ij}.
\end{eqnarray}

In summary, the TD-OMP2 method is defined by the EOMs
of double excitation amplitudes $\tau^{ab}_{ij}$ [Eq.~(\ref{t2eqn1})],
the relation $\lambda^{ij}_{ab} = \tau^{ab*}_{ij}$, and the EOMs of orbitals
[Eqs.~(\ref{orbeom1}) and (\ref{orbeom2})] with 1RDM and 2RDM elements given by
Eqs.~(\ref{eq:td-omp2_1rdm})-(\ref{eq:td-omp2_2rdm_corr}).
The orbital time-derivative terms $-iX$ can be dropped from
Eq.~(\ref{eq:fockix}), if one makes an arbitrary choice of
$\langle\psi_i|\dot{\psi_j}\rangle = \langle\psi_a|\dot{\psi_b}\rangle =0$ for 
the redundant orbital rotations.
It should be noted that (i) our partitioning scheme [Eqs.~(\ref{eqs:part})] is 
consistent with the standard M{\o}ller-Plesset perturbation theory in the 
absence of the external field $V_{\text{ext}}$, and 
(ii) in case with $V_{\text{ext}}$, the zeroth-order 
Hamiltonian is time dependent, both through the change of orbitals and 
an explicit time dependence of $V_{\text{ext}}(t)$, 
the latter implying nonperturbative inclusion of the laser-electron interaction. 

\subsection{Alternative derivation}
Another, more perturbation theoretic derivation begins with the 
following Lagrangian, 
\begin{eqnarray}\label{eq:td-omp2_lag2}
L = L_0 -i\lambda^{ij}_{ab}\dot{\tau}^{ab}_{ij}+&&
\langle
\Phi|\{(1+\hat{T}^\dagger_2)(\hat{f}_N-i\hat{X})(1+\hat{T}_2)
\nonumber \\
&&\hspace{3em}+
\hat{T}^\dagger_2\hat{v}_N + \hat{v}_N\hat{T}_2\}|\Phi\rangle_c,
\end{eqnarray}
where the subscript $c$ implies restriction to connected terms, and
$\hat{f}_N$ and $\hat{v}_N$ are the normal-ordered part of $\hat{f}$ and $\hat{v}$,
respectively. Then we follow the procedure of time-dependent variational principle
based on the action of Eq.~(\ref{eq:action}), to obtain identical method as derived
in the previous section. 
%
Two expressions of the TD-OMP2 Lagrangian, 
Eqs.~(\ref{eq:td-omp2_lag}) and (\ref{eq:td-omp2_lag2}), take the same numerical value as a function of time when evaluated with the solution of TD-OMP2, $\{\tau^{ab}_{ij}(t), \psi_p(t)\}$.
The Lagrangian of Eq.~(\ref{eq:td-omp2_lag2}) can be viewed as the time-dependent
extension of the Hylleraas energy functional used in conventional{\color{black}, time-independent} OMP2 method.
\cite{adamowicz1987optimized, neese2009assessment}

\section{Numerical results and discussion}\label{sec3}
\subsection{Ground-state energy of BH}
To assess the performance of the method described in the previous section, we do a series of the ground-state energy calculations taking BH molecule as an example
with double-$\zeta$ plus polarization (DZP) basis. \cite{HARRISON1983386} 
The imaginary time relaxation method\cite{sato2018communication} is used 
to obtain the ground state.
We have started our calculations with the bond length of 1.8 a.u. and gradually increased to 7.0 a.u., beyond which we could not achieve convergence.
The values for both the MP2 and OMP2 are reported in Table \ref{tab:comparison}. The required matrix elements are obtained from the Gaussian09 program \cite{gaussian09} package and 
interfaced with our numerical code.
All the values are compared with the values from the PSI4 program package. \cite{psi4}
We obtained identical results for the reported values except in five cases, even for which the difference appears only at the eighth or ninth digit 
after the decimal point. 
{\color{black} In these calculations, the number of orbitals $N$ are taken to be the same as the number of basis $n_{\text{bas}}$
to make a comparison with PSI4, \cite{psi4} and all the orbitals are treated as active. 
However, our implementation allows to optimize orbitals with $N \leq n_{\text{bas}}$ in general, 
with a flexible classification of the occupied orbital space into frozen core, dynamical core, and active.}

\begin{figure}[!t]
\centering
\includegraphics[width=1.0\linewidth]{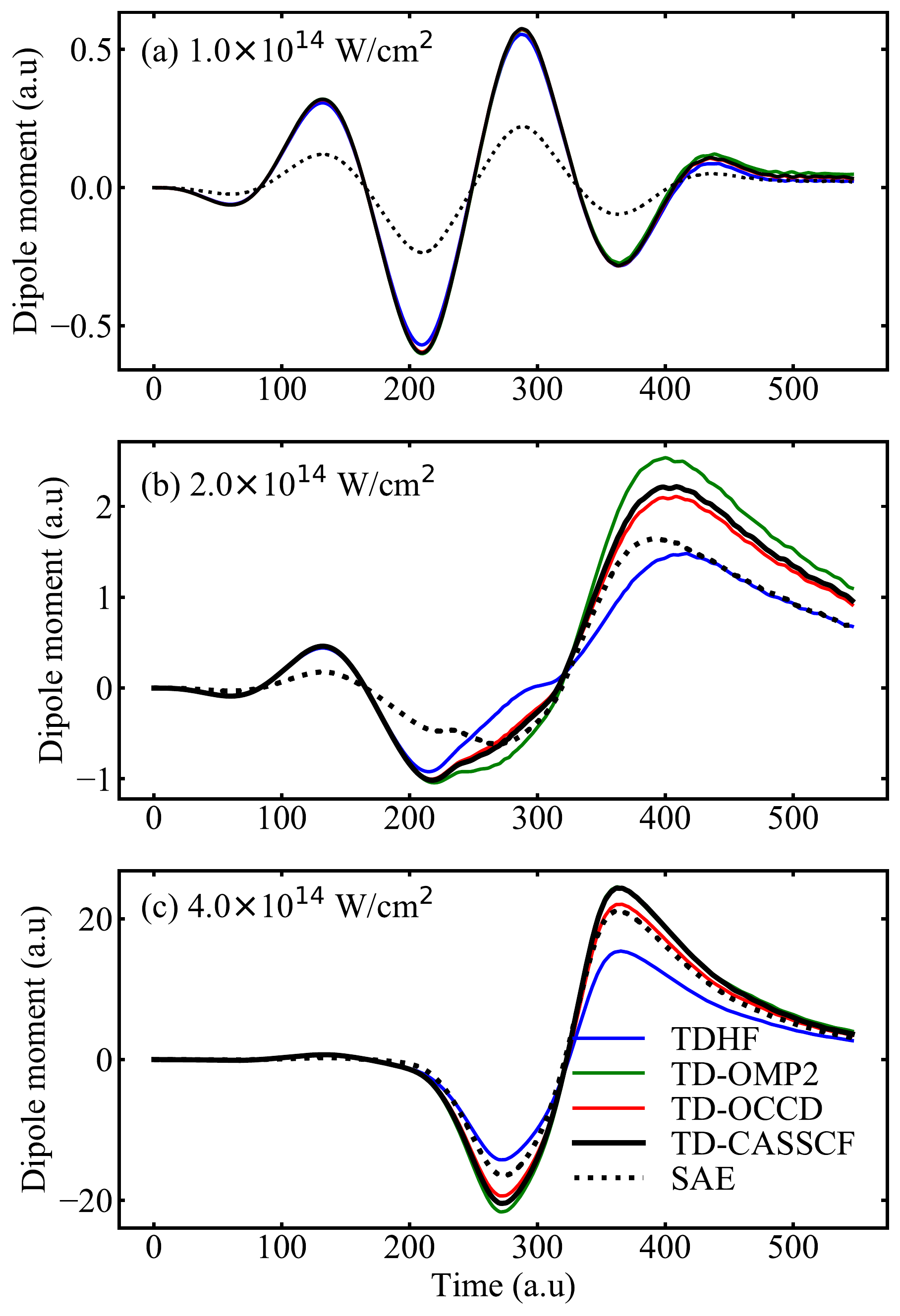}
\caption{\label{fig:ardipole}
{\color{black} Time evolution of {\color{black} sign-flipped} dipole moment {\color{black} $\langle z \rangle$} of Ar irradiated by
a laser pulse with a wavelength of 1200 nm and a peak intensity of 1$\times$10$^{14}$ W/cm$^2$ (a),
2$\times$10$^{14}$ W/cm$^2$ (b) and 4$\times$10$^{14}$ W/cm$^2$ (c),
calculated with TDHF, TD-OMP2, TD-OCCD, TD-CASSCF, {\color{black}and SAE}
methods.}
}
\end{figure}

{\subsection{Ar in a strong laser field}

We present numerical applications of the present method to Ar subject to an intense laser pulse linearly polarized in the $z$ direction.
Calculations for intense long-wavelength laser fields are computationally demanding, thus serving as a good test for the newly implemented methods.
The laser-electron interaction is introduced to the one-body part of the electronic Hamiltonian
within the dipole approximation in the velocity gauge,
\begin{eqnarray}\label{eq:h1vg}
h(\bm{r},\bm{p})=\frac{1}{2}|\bm{p}|^2 - \frac{Z}{|\bm{r}|} + A(t)p_z, 
\end{eqnarray}
where $Z(=18)$ is the atomic number, $A(t)=-\int^t E(t^\prime) dt^\prime$ is
the vector potential, with $E(t)$ being the laser electric field.
While our method is gauge-invariant, we obtain faster convergence with the velocity gauge for the case of intense long-wavelength laser pulses \cite{sato2016time,orimo2018implementation}.

\begin{figure}[!t]
\centering
\includegraphics[width=1.0\linewidth]{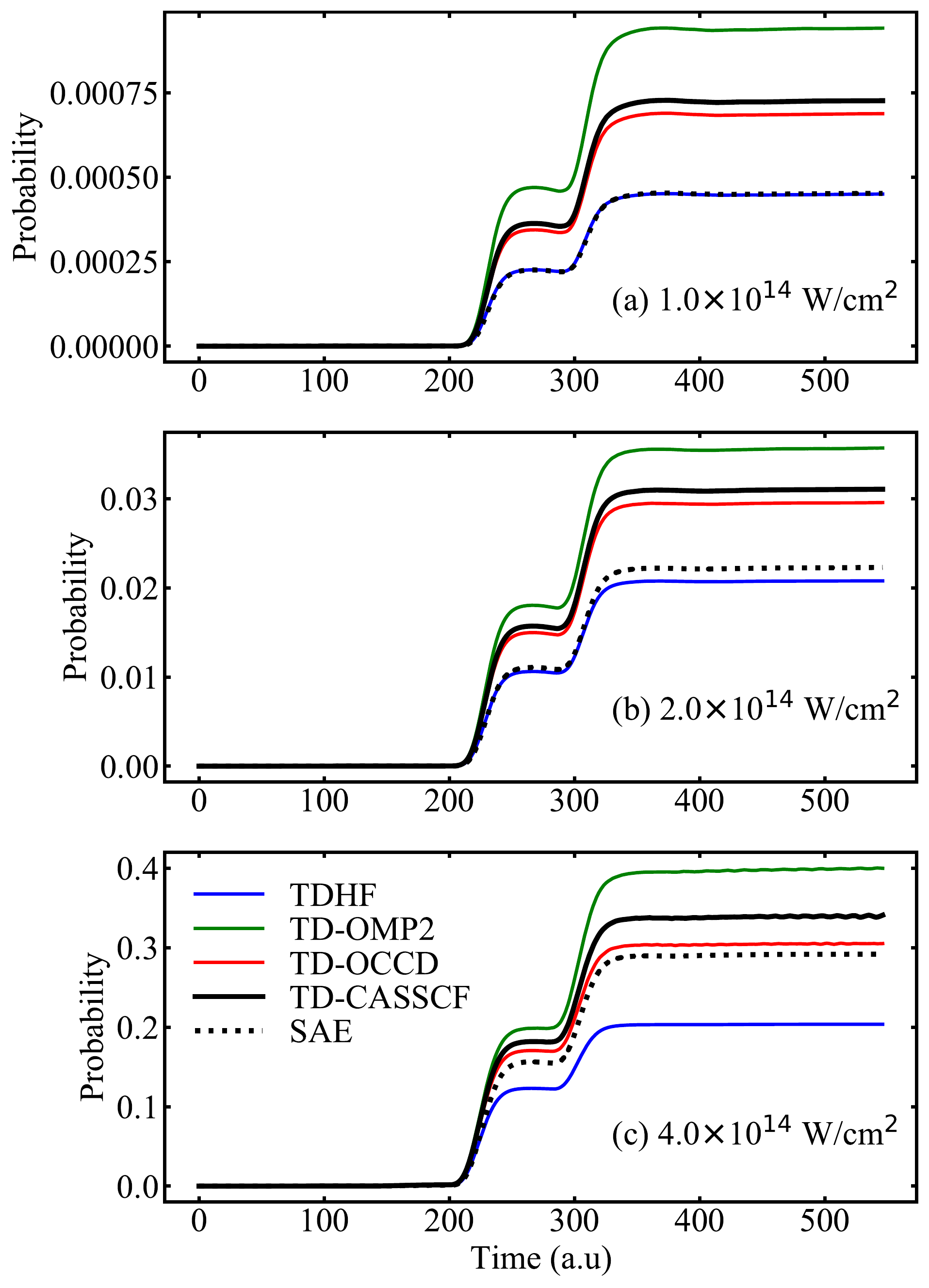}
\caption{\label{fig:arsip}
{\color{black} Time evolution of single ionization probability of Ar irradiated by
a laser pulse with a wavelength of 1200 nm and a peak intensity of 1$\times$10$^{14}$ W/cm$^2$ (a),
2$\times$10$^{14}$ W/cm$^2$ (b) and 4$\times$10$^{14}$ W/cm$^2$ (c)
calculated with TDHF, TD-OMP2, TD-OCCD, TD-CASSCF, {\color{black}and SAE}
methods.}
}
\end{figure}

The laser electric field is of the following form
\begin{eqnarray}
E(t)=E_0\,{\text{sin}}(\omega_0t)\,{\text{sin}}^2\left(\pi\frac{t}{3T}\right),
\end{eqnarray}
{\color{black}
for $0\le t\le 3T$, and $E(T)=0$ otherwise, with the central wavelength
$\lambda = 2\pi/\omega_0 = 1200$ nm, the period $T = 2\pi/\omega_0 \sim 4.00$ fs, 
and the peak intensity $I_0=E_0^2$.}
We have considered three different intensities of 1, 2, and 
4 $\times$10$^{14}$ W/cm$^2$ for Ar.

The spherical finite-element discrete variable representation (FEDVR) basis \cite{sato2016time, orimo2018implementation}
is used in our implementation. The convergence with respect to the
maximum angular momentum $l_{\text{max}}$ is checked at the TDHF level, and
$l_{\text{max}}$ is set to 72. {\color{black}The radial coordinate $r$ is
discretized by FEDVR consisting of 78 finite elements with 23 DVR
functions each, to support $0 < r < R_{\text{max}} = 300$.}
A ${\cos^\frac{1}{4}}$ mask function is switched on at 240 to avoid reflection
from the boundary. Fourth-order exponential Runge-Kutta integrator \cite{exponential_integrator} is used to propagate
equations of motions with 20000 time steps per optical cycle. The simulations are continued after the end of 
pulse {\color{black} for further 6000 time steps}}.

For {\it ab initio} TDHF, TD-OMP2, TD-OCCD, and TD-CASSCF methods,
the $1s2s2p$ core is kept frozen, and the dynamics of remaining eight electrons are 
actively taken into account with four (TDHF) or thirteen (TD-OMP2, TD-OCCD, and TD-CASSCF) active orbitals.
{\color{black}
The SAE method first diagonalizes the following effective Hamiltonian\cite{Muller:1998,Schiessl:2006}
\begin{eqnarray}\label{eq:sae}
h_{\rm eff} = \frac{1}{2}|\bm{p}|^2+V_{\rm eff}(|\bm{r}|),
\end{eqnarray}
on the FEDVR basis, where the effective potential $V_{\rm eff}(r)$ is taken to be\cite{Muller:1998,Schiessl:2006}
\begin{eqnarray}
V_{\rm eff}(r) = -\frac{1}{r}\left\{
1+Ae^{-r}+(Z-1-A)e^{-Br}
\right\}, 
\end{eqnarray}
with $A=5.4$ and $B=3.682$ for Ar \cite{Schiessl:2006}. This potential correctly supports
$1s$, $2s$, $2p$, $3s$, and $3p$ bound orbitals, with the $3p$ orbital energy of -15.82 eV with the present FEDVR basis. 
After obtaining the ground state, we solve the effective one-electron Schr{\"o}dinger equation, 
\begin{eqnarray}\label{eq:sae}
i\frac{d}{dt}|\chi\rangle = 
\hat{Q} \left\{h_{\rm eff} + A(t)p_z\right\} |\chi\rangle,
\end{eqnarray}
starting from the $3p_0$ orbital. The projector $\hat{Q}=1-\sum_j |\phi_j\rangle \langle\phi_j|$, with $\phi_j$ running over
$1s$, $2s$, $2p$, $3s$, and $3p_{\pm}$ orbitals [multiplied by the gauge factor $e^{-iA(t)z}$], keeps $\chi(t)$ orthonormal to
the inner shell orbitals.
}

In Fig.~\ref{fig:ardipole}, we plot the time evolution of the {\color{black} sign-flipped dipole moment $\langle z\rangle$}
evaluated as a trace $\langle z\rangle = \langle \psi_p|\hat z |\psi_q\rangle \rho^q_p$ 
{\color{black}for {\it ab initio} methods and $\langle z\rangle = \langle\chi|z|\chi\rangle$ for SAE}.
We compare the TD-OMP2 results with {\color{black}SAE,} TDHF, TD-OCCD, and TD-CASSCF ones.
Within the same active space, TD-CASSCF produces highly accurate results, useful for performance analysis of the TD-OMP2 method.

{\color{black}
The lowest ($1.0\times 10^{14}$ W/cm$^2$) and highest ($4.0\times 10^{14}$ W/cm$^2$) intensity cases
characterize the dynamics with small and substantial amount of ionization, respectively (Fig.~\ref{fig:arsip} below). The SAE approximation is known\cite{krause1992jl, kulander1987time,Muller:1998}
to work better for the latter case, 
where the dynamics is dominated by tunneling ionization of the single, most weakly bound electron (one of the $3p_0$ electrons in the present case).
It is not well suited for describing the former case dominated by collective bound dynamics. On the other hand, TDHF serves as the
reference multielectron method without the (Coulomb) correlation by definition, and provides a qualitatively correct description of the bound dynamics.
It, however, fails to accurately describe cases with sizable tunneling ionization. (See, e.g, Ref.~\citenum{Sato:2014} and references therein.)
These trends of SAE and TDHF methods are well confirmed in the performance comparison to the reference TD-CASSCF method in Figs.~\ref{fig:ardipole} and \ref{fig:arsip}.
}

\begin{figure}[!b]
\centering
\includegraphics[width=1.0\linewidth]{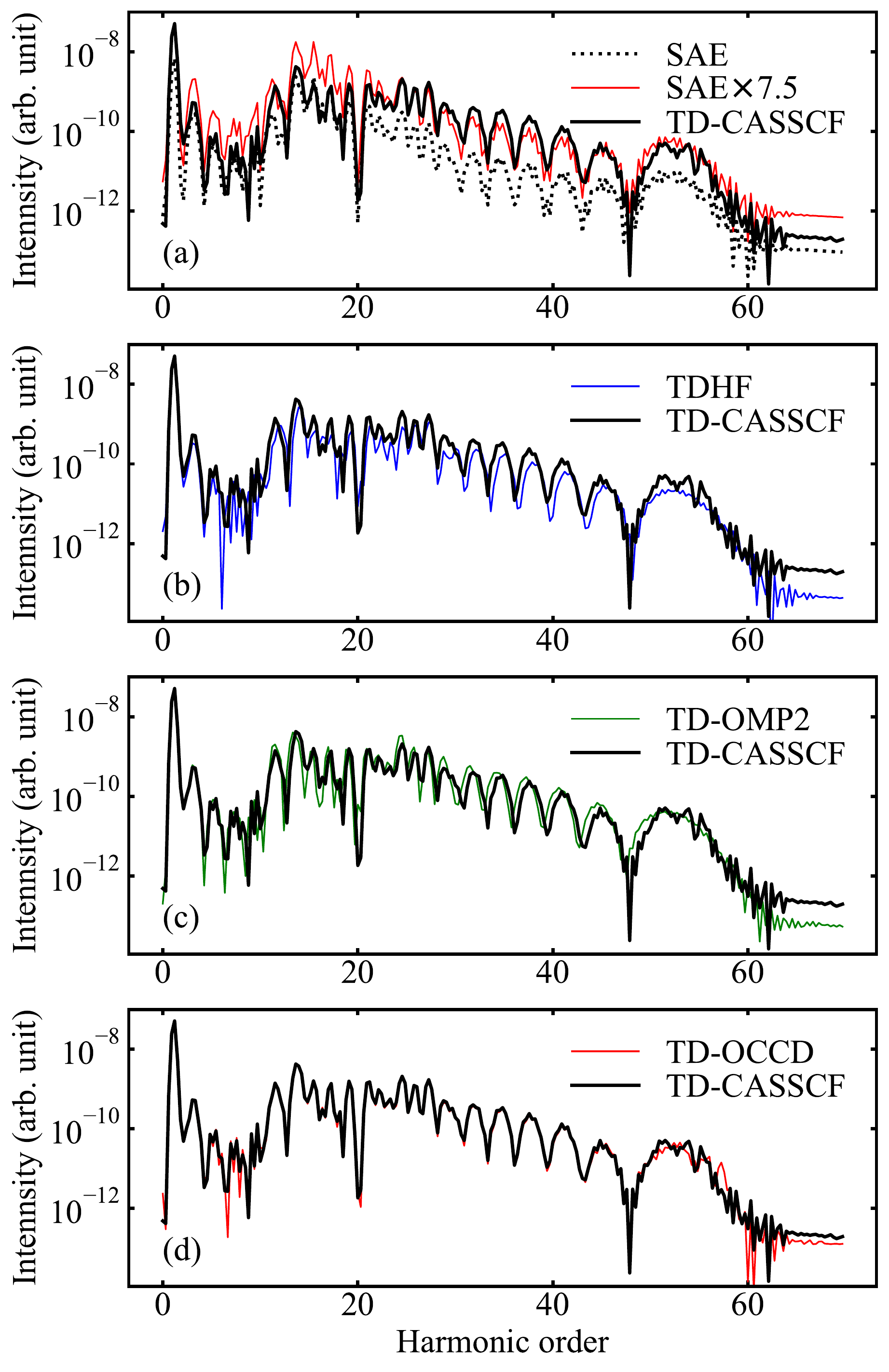}
\caption{\label{fig:arhhg1}
{\color{black}
The HHG spectra from Ar irradiated by a laser pulse
with a wavelength of 1200 nm and 
a peak intensity of 1$\times$10$^{14}$ W/cm$^2$.
Comparison of the results of SAE (a), TDHF (b), TD-OMP2 (c), and TD-OCCD (d) methods with that of TD-CASSCF.
}}
\end{figure}

\begin{figure}[!t]
\centering
\includegraphics[width=1.0\linewidth]{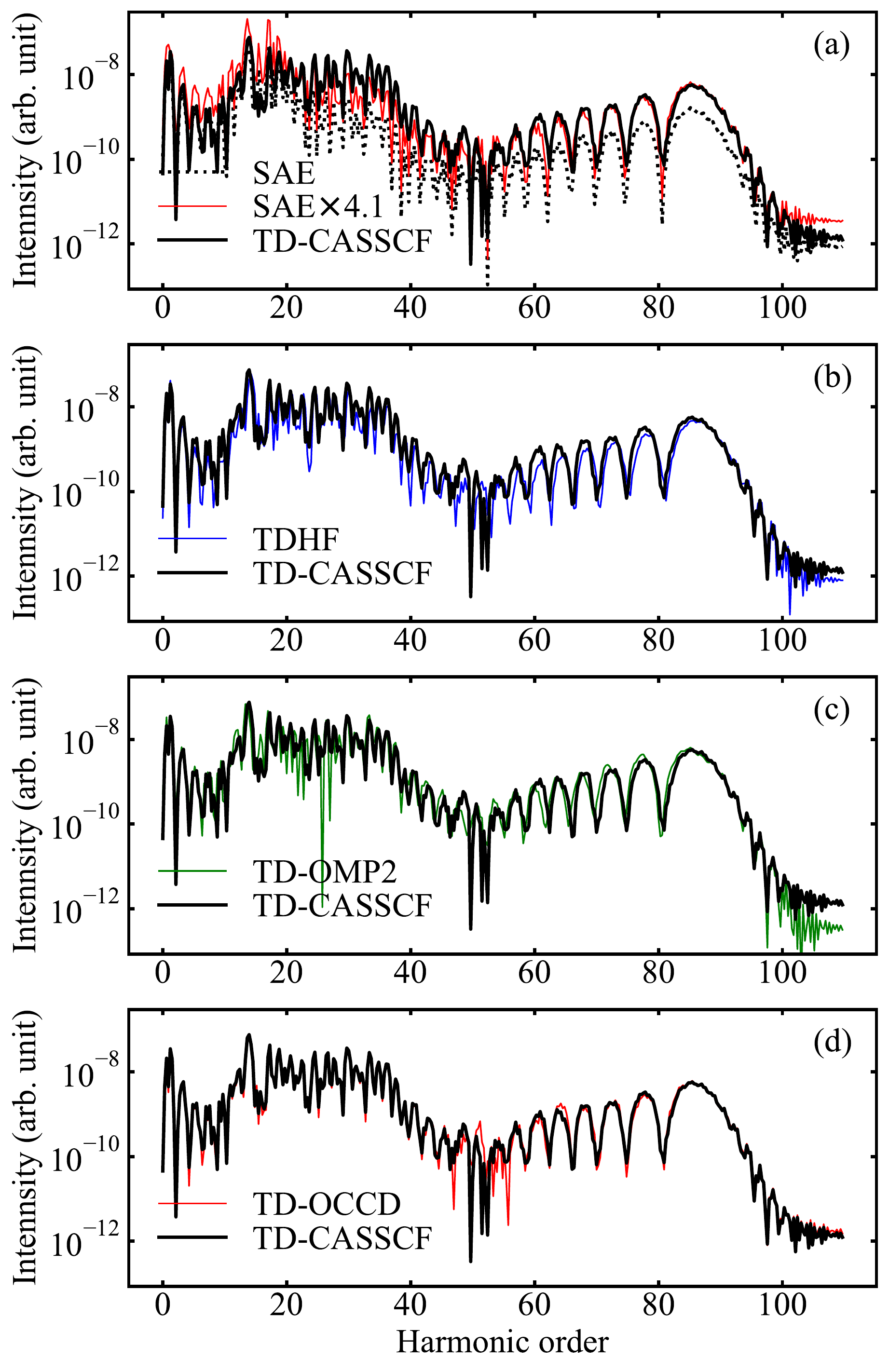}
\caption{\label{fig:arhhg2}
{\color{black}
The HHG spectra from Ar irradiated by a laser pulse
with a wavelength of 1200 nm and 
a peak intensity of 2$\times$10$^{14}$ W/cm$^2$.
Comparison of the results of SAE (a), TDHF (b), TD-OMP2 (c), and TD-OCCD (d) methods with that of TD-CASSCF.
}}
\end{figure}

\begin{figure}[!t]
\centering
\includegraphics[width=1.0\linewidth]{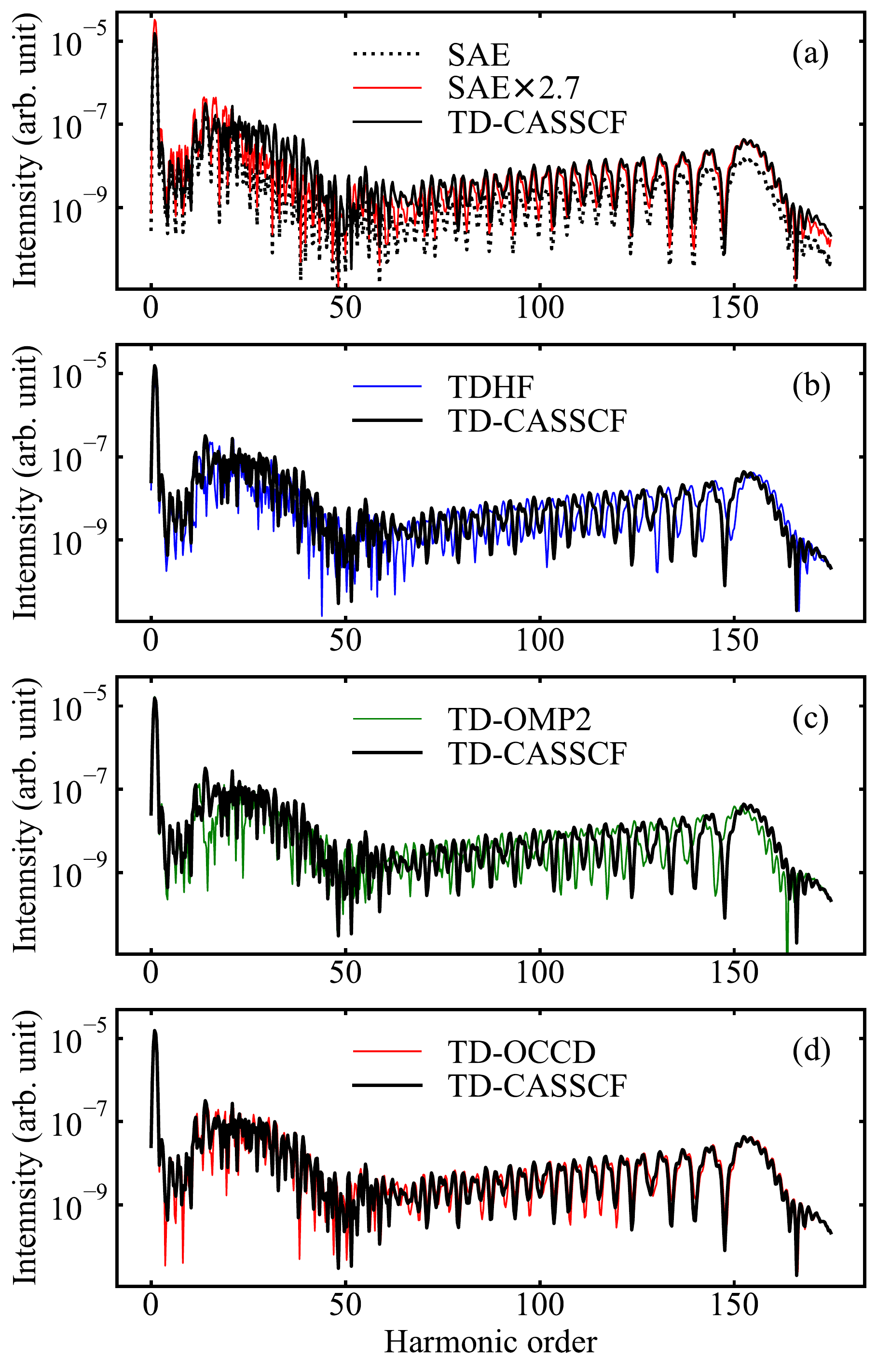}
\caption{\label{fig:arhhg4}
{\color{black}
The HHG spectra from Ar irradiated by a laser pulse
with a wavelength of 1200 nm and 
a peak intensity of 4$\times$10$^{14}$ W/cm$^2$.
Comparison of the results of SAE (a), TDHF (b), TD-OMP2 (c), and TD-OCCD (d) methods with that of TD-CASSCF.
}}
\end{figure}

As seen in Fig.~\ref{fig:ardipole}-(a), at the lowest intensity of $10^{14}\,{\rm W/cm}^2$ all the methods {\color{black}except for SAE} produce similar results.
The TD-OCCD produces virtually the identical result with the TD-CASSCF, whereas TD-OMP2 slightly overestimates, and TDHF underestimates,
considering TD-CASSCF as the benchmark.
With increase in intensity, the difference among the methods become more prominent.
While all the methods {\color{black}except for SAE} give similar results in the early stage, TD-OMP2 and TDHF start to overestimate and underestimate, respectively, with the progress of tunneling ionization (Fig.~\ref{fig:arsip}).
In general, the performance of the TD-OMP2 method is better than TDHF {\color{black}and SAE} due to consideration at least a part of the electron correlation.
{\color{black}
It is noticed that the TD-OMP2 dipole moment agrees better with the TD-CASSCF one
for the highest intensity [Fig.~\ref{fig:ardipole}-(c)] than for the
intermediate one [Fig.~\ref{fig:ardipole}-(b)], which might indicate that
the latter case, with both sizable ionization and nontrivial correlation effects coexisting,
is theoretically more challenging.}

The general trends in Fig.~\ref{fig:ardipole} are also found in the single ionization probability (Fig. \ref{fig:arsip}), 
evaluated as the probability of finding an electron outside a sphere of 20 a.u.~radius.
Again, we see a systematic overestimation by TD-OMP2 and underestimation by TDHF in comparison to the TD-CASSCF result; the performance of TD-OMP2 is better than that of TDHF
{\color{black}and SAE, except for the highest intensity case, where the SAE result is as accurate
as the TD-OCCD one}.

\begin{figure}[!t]
\centering
\includegraphics[width=1.0\linewidth]{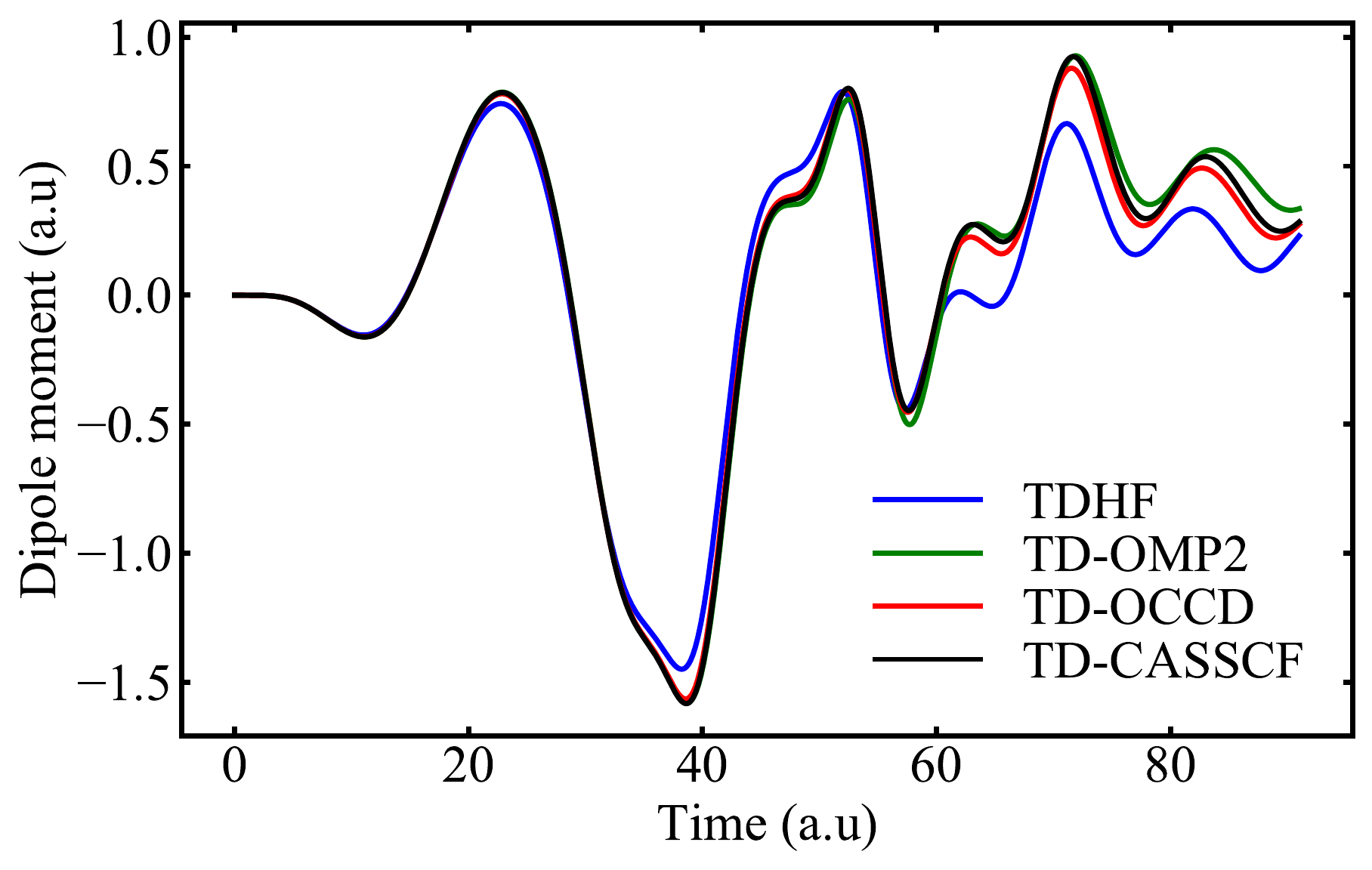}
\caption{\label{fig:ar200nm}
{\color{black} Time evolution of sign-flipped dipole moment $\langle z \rangle$ of Ar irradiated by
a laser pulse with a wavelength of 200 nm and a peak intensity of 4$\times$10$^{14}$ W/cm$^2$,
calculated with TDHF, TD-OMP2, TD-OCCD, and TD-CASSCF methods.}}
\end{figure}

\begin{table}[!b]
\caption{\label{tab:timing_electron} \color{black} Comparison of the CPU                                                             
time$^\text{a}$ (in second) spent for the evaluation of the $T_2$ equation,                                                                     
$\Lambda_2$ equation, and 2RDM for TD-OCCD and TD-OMP2 methods}
\begin{ruledtabular}
\begin{center}
\begin{tabular}{rrrrrrr}
\multicolumn{3}{c}{TD-OCCD} && \multicolumn{3}{c}{TD-OMP2}\\
\cline{1-3} \cline{5-7}
\multicolumn{1}{c}{T$_2$}&\multicolumn{1}{c}{$\Lambda_2$}&\multicolumn{1}{c}{2RDM}&
&\multicolumn{1}{c}{T$_2$}&\multicolumn{1}{c}{$\Lambda_2$}&\multicolumn{1}{c}{2RDM}\\
\hline \\
\, 40.8&\, 55.5&\, 109.5&\,&\, 1.11&\,-&\,0.25\\
\end{tabular}
\end{center}
\end{ruledtabular}
\footnotetext[1]                                                                                                                     
{\color{black} CPU time spent for the simulation of Ar atom for                                                                      
 1000 time steps ($0 \leq t \leq 0.05T$) of a real-time simulation                                                                                           
 ($I_0=4\times 10^{14}$ W/cm$^{2}$ and $\lambda=1200$ nm.),
 using an Intel(R) Xeon(R) CPU with 12 processors having a clock speed of 3.33GHz.}
\end{table}

In Figs.~\ref{fig:arhhg1}--\ref{fig:arhhg4} we compare HHG spectra, calculated as the modulus squared of the Fourier transform of the expectation value of the dipole acceleration, which, in turn, is obtained with a modified Ehrenfest expression\cite{sato2016time}.
All {\it\color{black} ab initio} methods reproduce the HHG spectra relatively well, including an experimentally observed characteristic dip around the 52nd order ($\sim 54$ eV) at $2\times$ and $4\times 10^{14}\,{\rm W/cm}^2$ related to the
Cooper minimum of the photoionization spectrum{\cite{PhysRevLett.102.103901} at the same energy.
However, TDHF systematically underestimates and fails to reproduce fine details.
{\color{black}The SAE method severely underestimates the HHG yields, although the overall shape of the HHG spectrum is well reproduced, and an intensity-dependent scaling brings the spectral intensity at the higher plateau close to that of TD-CASSCF, especially for the highest intensity.}
The agreement with the TD-CASSCF results is much better for the TD-OCCD method, which contains nonlinear terms in the amplitude equations, then followed by TD-OMP2 with slight overestimation. 
This trend is consistent with the capability of each method to treat the electron correlation.

{\color{black} 
In order to investigate the performance of TD-OMP2 method for shorter wavelengths,
where electrons not only in the highest-occupied but also in the inner-shell orbitals are
driven by the laser, we consider a shorter wavelength of 200 nm with an intensity of 4$\times$10$^{14}$ W/cm$^2$.
All the other simulation parameters are identical to those given above. 
The obtained time evolution of the dipole moment (Fig.~\ref{fig:ar200nm}) shows
a slight overestimation of the oscillation amplitude by TD-OMP2 and underestimation
by TDHF compared to the TD-CASSCF result as in the case for the longer wavelength.
It is encouraging, however, that the TD-OMP2 result is much closer to the
TD-CASSCF one for the shorter wavelength, which is more sensitive to the treatment of correlation effects.
}

It is worth noting that, for an intensity as high as $4\times 10^{14}\,{\rm W/cm}^2$, 
the TD-OMP2 {\color{black} simulation completed stably.
This should be the direct consequence of full inclusion of the laser-electron interaction 
in the zeroth-order Hamiltonian and the orbital optimization (propagation) according to
the time-dependent variational principle based on the total (up to second-order) Lagrangian, 
which keeps the instantaneous {\it perturbation} $\hat{H}^{(1)} = \hat{H}-\hat{f}$ 
relatively small in the present simulation.}

{\color{black}
Finally, the computational cost of different parts of TD-OMP2 and TD-OCCD methods are compared in 
Table \ref{tab:timing_electron} for the same computational condition as in Fig.~\ref{fig:arhhg1}~(c).
The evaluation of $T_2$ equation, $\Lambda_2$ equation, and 2RDM all
scale as $N^6$ for the TD-OCCD method.
On the other hand, {\color{black}for the TD-OMP2 method,} the evaluation of $T_2$ equation scales as $N^5$, and
we do not need a separate solution for $\Lambda_2$, as it is just the
complex conjugate of $T_2$. The greatest time saving for the TD-OMP2
method comes from the evaluation of 2RDM since it scales as $N^4$,
and does not involve any operator products.
Overall, TD-OMP2 achieves a significant cost reduction compared to TD-OCCD,
making it an attractive choice for simulations involving
larger chemical systems.}

\section{conclusion}\label{sec4}
We have successfully implemented the TD-OMP2 method for the real-time simulations of laser-induced dynamics in relatively large chemical systems.
The TD-OMP2 method retains the size-extensivity and gauge-invariance of TD-OCC, and is computationally much more efficient than the full TD-OCCD method.
As a first numerical test, we have applied the method to the ground state of BH and the laser-driven dynamics of Ar.
The imaginary time relaxation for BH obtains the identical ground-state energies with those by the stationary theory, 
which indicates the correctness of the implementation.
The performance of the present method is {\color{black}
numerically investigated in comparison to SAE, TDHF, 
TD-OCCD, and TD-CASSCF methods}
for the case of laser-driven Ar. 
The results suggest a decent performance with a consistent overestimation of the correlation effect in such highly nonlinear phenomena.
Remarkably, the TD-OMP2 method is stable and does not 
breakdown even {\color{black}in the presence of strong laser-electron interaction, 
thanks to the nonperturbative inclusion of external fields and time-dependent orbital optimization. Further assessment of the TD-OMP2 method for different systems and severer simulation 
conditions (e.g, with higher-intensity and/or longer-wavelength lasers) will be reported 
elsewhere.}

\section*{acknowledgments}
{\color{black}
This research was supported in part by a Grant-in-Aid for
Scientific Research (Grants No. 17K05070,
No. 18H03891, and No. 19H00869) from the Ministry of Education, Culture,
Sports, Science and Technology (MEXT) of Japan. 
This research was also partially supported by JST COI (Grant No.~JPMJCE1313), JST CREST (Grant No.~JPMJCR15N1),
and by MEXT Quantum Leap Flagship Program (MEXT Q-LEAP) Grant Number JPMXS0118067246.}

\section*{DATA AVAILABLITY}
The data that support the findings of this study are available from the corresponding author
upon reasonable request.
\bibliographystyle{h-physrev}
\bibliography{ref}
\end{document}